\documentclass[aps,prl,superscriptaddress,showpacs,longbibliography,floatfix,twocolumn]{revtex4-1}

\usepackage[utf8]{inputenc}
\usepackage{slashed}
\pdfoutput=1

\usepackage{color}
\usepackage{graphicx}   
\usepackage{bm}
\usepackage{amsmath}
\usepackage{amsfonts}
\usepackage{hyperref}

\def\be{\begin{equation}}
\def\ee{\end{equation}}
\def\ba{\begin{eqnarray}}
\def\ea{\end{eqnarray}}

\begin{document}

\title{What if $\phi^4$ theory in 4 dimensions is non-trivial in the continuum?}

\author{Paul Romatschke}
\affiliation{Department of Physics, University of Colorado, Boulder, Colorado 80309, USA}
\affiliation{Center for Theory of Quantum Matter, University of Colorado, Boulder, Colorado 80309, USA}

\begin{abstract}
  Traditionally, scalar $\phi^4$ theory in four dimensions is thought to be quantum trivial in the continuum. This tradition is apparently well grounded both in physics arguments and mathematical proofs. Digging into the proofs one finds that they do not actually cover all physically meaningful situations, in particular the case of multi-component fields and non-polynomial action. In this work, I study multi-component scalar field theories in four dimensions in the continuum and show that they do evade the apparently foregone conclusion of triviality. Instead,  one finds a non-trivial interacting theory that has two phases, bound states and non-trivial scattering amplitudes in the limit of many components. This has potentially broad implications, both for the foundations of quantum field theory as well as for the experimentally accessible Higgs sector of the Standard Model.
\end{abstract}

\maketitle

It is commonly thought that scalar field theory with quartic interaction is quantum trivial in the continuum, meaning that all $n$-point functions are Gaussian, or equivalently, that the renormalized coupling vanishes in the infrared.

The notion of quantum triviality dates back at least to Wilson's application of the renormalization group (cf. Ref.~\cite{Wilson:1973jj} for a review). In a nutshell, a version of the physics story for quantum triviality in scalar $\phi^4$ theory in four dimensions goes like this: renormalizing the theory, we find a positive $\beta$-function, so the renormalized coupling always increases. Because of this, the renormalized coupling in the UV must be bigger than in the IR. We also want a UV fixed point for the theory, where the coupling is finite. In the continuum, we need to push the UV-cutoff to infinity, and since the coupling increases, the renormalized coupling in the IR must vanish.

The renormalization group argument suffers from the problem that the $\beta$-function in $\phi^4$ theory can usually only be calculated in perturbation theory, precluding statements about the interesting regime where the running coupling becomes large. In contrast, mathematical proofs of quantum triviality in $\phi^4$ theory such as those in Refs.~\cite{Frohlich:1982tw,Aizenman:2019yuo} use a mapping of field theory to the Ising model which does not rely on perturbative expansions.

However, it should be recalled that the mathematical proofs of triviality are based on specific assumptions, namely that the scalar field has only one or two components (not more), that the interaction is polynomial and that the bare coupling constant is positive.

While these seem innocent enough, they do not cover the situations considered in this work.

In particular, I will consider the case of many scalar fields (also known as the O(N) model), because this allows for controlled calculations outside the usual perturbative regime.
Also, I will allow for the possibility that the coupling constant is non-positive. Naively, this seems to violate all requirements of unitarity and positivity that are associated with ``proper'' quantum mechanics. However, in a remarkable work Bender and B\"ottcher \cite{Bender:1998ke} were able to show that non-positive couplings do lead to positive-definite Hamiltonian spectra in quantum mechanics in what has become known as ${\cal PT}$-symmetric quantum mechanics. In fact, for non-positive, but ${\cal PT}$-symmetric Hamiltonians with quartic interaction, positivity has been proved \cite{Dorey:2001uw} by noting that these naively unbounded Hamiltonians are spectrally equivalent to ordinary (positive-definite) Hamiltonians.

Application of ${\cal PT}$-symmetry to quantum field theory is still in its infancy \cite{Bender:2021fxa,Mavromatos:2021hpe,Grunwald:2022kts,Ai:2022csx}, but encouraging results for non-positive coupling  in the case of N-component fields exist \cite{Romatschke:2022jqg,Grable:2023paf,Lawrence:2023woz}. On a technical level, this is brought about by contour-deforming the integral domain of the path integral from the real to the complex domain, similar to how analytic continuation of e.g. the Gamma-function is introduced in mathematics.

For concreteness, I will provide calculations in the large N limit of the O(N) model, corresponding to many copies of the scalar field interacting with a quartic potential. This theory has all the signs of being quantum trivial according to the usual beliefs: it has strictly positive $\beta$-function and a Landau pole \cite{Romatschke:2022jqg,Romatschke:2022llf}. However, this theory can be studied without using perturbation theory in the large N limit, and thus evades the restriction to the small coupling regime.

Rumor has it that the presence of a Landau pole, or a point where the coupling constant diverges, is a fatal flaw of continuum quantum field theory. How can a theory with a divergent coupling be considered physical? To address this criticism preemptively, let me remind readers that the renormalized coupling constant is in general not a physical observable in four-dimensional quantum field theory. Instead, it enters physical observables in often quite complicated fashion, resulting in perfectly finite and reasonable results for these observables even when the coupling diverges\footnote{For a very incomplete list of examples in various field theories, cf. Refs.~\cite{Parisi:1975im,Itzhaki:1998dd,Klebanov:2002ja,Romatschke:2019ybu,Romatschke:2021imm}.}. I will present a detailed calculation showing that this is indeed what happens for the O(N) model in the main part of this work.

For good measure, I will also study one-component $\phi^4$ theory with negative  coupling on the lattice, using contour deformation, in a modern twist of Symanzik's proposal a little while ago \cite{Symanzik:1973hx}.

\section{Calculation}

I consider an N-component scalar field $\vec{\phi}=\left(\phi_1,\phi_2,\ldots,\phi_N\right)$ in $d$ Euclidean dimensions with the Euclidean Lagrangian density
\be
{\cal L}_E=\frac{1}{2}\partial_\mu \vec{\phi}\cdot \partial_\mu \vec{\phi}+\frac{1}{2}m_{0}^2 \vec{\phi}^2+\frac{\lambda_{0}}{N}\left(\vec{\phi}\cdot \vec{\phi}\right)^2\,,
\ee
where $m_0$ is the bare mass parameter and $\lambda_0$ is the bare coupling constant. The partition function for this theory is given by the path integral over the fields,
\be
\label{Z1}
Z=\int {\cal D}\vec{\phi} e^{-S_E}\,.
\ee

If find it useful to perform a Hubbart-Stratonovic transformation of the path integral by inserting a path integral over the auxiliary field $\zeta$ as
\be
e^{-\int dx \frac{\lambda_0}{N}\left(\vec{\phi}\cdot \vec{\phi}\right)^2}=\int {\cal D}\zeta e^{- \int dx \left[\frac{i}{2} \zeta \vec{\phi}^2+\frac{\zeta^2 N}{16 \lambda_0}\right]}\,.
\ee
As a consequence of this transformation, the path integral over the scalar field becomes Gaussian, and formally be solved exactly so that $Z=\int {\cal D}\zeta e^{-N A}$ with the action
\be
\label{action}
A\!=\!\frac{1}{2}{\rm tr}\ln \left[-\partial^2+m_0^2+i \zeta\right]\!+\!\int dx \frac{\zeta^2}{16\lambda_0}\,.
\ee

So far, everything has been exact, but since the remaining path integral cannot be solved in closed form, one has to choose an expansion scheme in order to make progress.

A common choice in quantum field theory is to resort to perturbation theory in the coupling constant $\lambda$. In fact, it seems that this choice is so common that sometimes researchers blur the distinction between quantum field theory and perturbative quantum field theory. However, for the present purpose, this is not a viable choice, because we will have to deal with situations where the coupling constant is not small, thereby invalidating the use of perturbation theory.

Fortunately, for the N-component scalar case, there is a different expansion scheme in the form of $\frac{1}{N}$ in the limit of $N\rightarrow \infty$. Since this expansion scheme does not involve the coupling constant, it is by construction non-perturbative, so rigorous statements outside the regime of perturbation theory are possible.

In the large N-limit, the path integral for the theory can be solved exactly by method of steepest descent, leading to
\be
\ln Z=-N A_{LO}+{\cal O}\left(N^0\right)\,,
\ee
where $A_{LO}$ is the action (\ref{action}) evaluated at the saddle point given by the condition $\frac{\delta A}{\delta \zeta(x)}=0$. It is well-known \cite{Moshe:2003xn,Romatschke:2019rjk} that the saddle point for this theory corresponds to a imaginary-valued constant $\zeta(x)=-i z^*$, with the saddle point condition taking the form
\be
\label{saddle}
\frac{z^*}{8\lambda_0}=\frac{1}{2}\int dk \frac{1}{k^2+m_0^2+z^*}\,.
\ee
It is easiest to deal with the momentum integral by using dimensional regularization in $d=4-2\varepsilon$ dimensions, but I've come to realize that a significant fraction of readers is mistrustful of dim-reg. For this reason, I regulate the momentum integral here using a UV cutoff $\Lambda_{\rm UV}$.

It is important to point out that if $\Lambda_{\rm UV}$ is kept finite, scalar $\phi^4$ theory becomes a so-called effective (or cut-off) theory. Most of the literature on this theory is based on this effective theory approach, and there is little controversy about effective $\phi^4$ theory being non-trivial. 

However, this is not the route taken here. Instead of keeping $\Lambda_{\rm UV}$ finite, I will take this cutoff to infinity after renormalizing the theory. This is necessary in order to study $\phi^4$ not as an effective cut-off theory, but as a bona-fide quantum field theory in the continuum.

In cut-off regularization, the saddle-point condition (\ref{saddle}) in four dimensions becomes
\be
\label{hu}
\frac{z^*}{\lambda_0}=\frac{1}{(2\pi)^2}\left[\Lambda_{\rm UV}^2+(m_0^2+z^*)\ln \frac{m_0^2+z^*}{\Lambda_{\rm UV}^2}\right]\,.
\ee
Recognizing from (\ref{action}) that the propagator mass involves the combination $m_0^2+z^*$, I add $\frac{m_0^2}{\lambda_0}$ to both sides of (\ref{hu}).

Renormalization of the saddle point condition then can be achieved by the non-perturbative renormalization conditions \cite{Moshe:2003xn}
\be
\label{renorm}
\frac{1}{\lambda_0}=\frac{1}{\lambda_R}+\frac{1}{(2\pi)^2}\ln \frac{\mu^2}{\Lambda_{\rm UV}^2}\,, \quad \frac{\Lambda_{\rm UV}^2}{(2\pi)^2}+\frac{m_0^2}{\lambda_0}=\frac{m_R^2}{\lambda_R}\,,
\ee
where $\lambda_R(\mu),m_R(\mu)$ are the renormalized coupling constant and mass parameter of the theory and $\mu$ is the (fictitious!) renormalization scale. Considering the massless (critical) theory with $m_R(\bar\mu)=0$, in terms of renormalized quantities, the saddle point condition becomes
\be
\frac{m_0^2+z^*}{\lambda_R}=\frac{m_0^2+z^*}{(2\pi)^2}\ln \frac{m_0^2+z^*}{\mu^2}\,,
\ee
which has two solutions $m_0^2+z^*=0$, $m_0^2+z^*=\mu^2 e^{\frac{(2\pi)^2}{\lambda_R(\mu)}}$. As intended, these solutions are finite as long as $\lambda_R$ is finite, which guarantees a finite pole mass for the propagator at leading order in the large N expansion.

\begin{figure}[t]
  \includegraphics[width=.95 \linewidth]{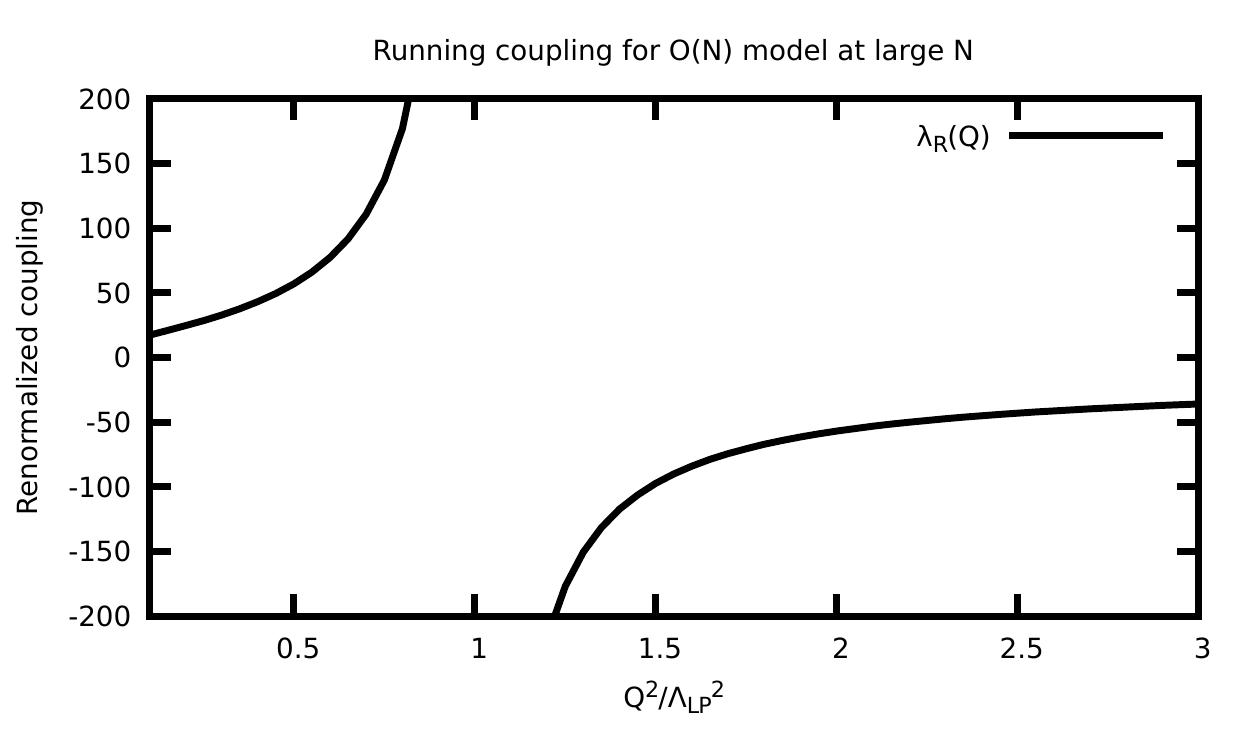}
  \caption{\label{fig1} Exact renormalized running coupling in the continuum limit of large N scalar $(\vec{\phi}^2)^2$ theory. See text for details.}
  \end{figure}

Again, it should be stressed that the calculation is based on a large N expansion scheme, so that in particular the renormalization conditions (\ref{renorm}) are valid at any value of the coupling, in particular in the non-perturbative regime. The renormalization condition for the coupling leads to the analytic form 
\be
\label{running}
\lambda_R(\mu)=\frac{1}{\frac{1}{\lambda_0}+\frac{1}{(2\pi)^2}\ln \frac{\Lambda_{\rm UV}^2}{\mu^2}}\,,
\ee
and the $\beta$-function
\be
\label{beta}
\beta=\frac{\partial \lambda_R(\mu)}{\partial \ln \mu^2}=\frac{\lambda_R^2(\mu)}{(2\pi)^2}\,.
\ee

The $\beta$-function is positive for all real-valued couplings $\lambda_R(\mu)$, suggesting that the running coupling $\lambda_R(\mu)$ always increases with increasing scale $\mu$. Also, the only fixed point where $\beta=0$ is at $\lambda_R(\mu)=0$. 

At the UV scale, $\lambda_R(\mu=\Lambda_{\rm UV})=\lambda_0=\lambda_{\rm ref}$. Fixing $\lambda_{\rm ref}>0$ to be some finite number at the UV scale, we may ask what happens to the coupling in the infrared, at some physics scale of interest $\mu^2=Q^2\neq 0$. However, in order to access the continuum theory, we still need to send $\Lambda_{\rm UV}\rightarrow \infty$. According to the exact running (\ref{running}), we thus have in the continuum limit
\be
\label{qtriv}
\lim_{\rm \Lambda_{\rm UV}\rightarrow \infty}\lambda_R(Q)=\frac{1}{\frac{1}{\lambda_{\rm ref}}+\frac{1}{(2\pi)^2}\ln \frac{\Lambda_{\rm UV}^2}{Q^2}}\rightarrow 0\,.
\ee

We seem to find that -- in the continuum limit -- the infrared properties of the theory are necessarily trivial, in the sense that coupling at any physically relevant scale is zero. In a nutshell, we seem to have proven quantum triviality for the N-component $\phi^4$ theory in the large N limit, in complete agreement with results for $N=1$ $\phi^4$ theory, cf. Refs.~\cite{Wilson:1973jj,Frohlich:1982tw,Aizenman:2019yuo} for a very incomplete list.

But now let's take a second look.

If we would like to keep the property that the coupling $\lambda_R(\mu)$ is unchanged as the cut-off is removed $\Lambda_{\rm UV}\rightarrow \infty$, this necessarily implies that
\be
\label{lambda0}
\lambda_0=\frac{(2\pi)^2}{\ln \frac{\Lambda_{\rm LP}^2}{\Lambda_{\rm UV}^2}}\,,
\ee
with $\Lambda_{\rm LP}$ a fixed momentum scale, such that
\be
\label{contrun}
\lim_{\Lambda_{\rm UV}\rightarrow \infty}\lambda_R(\mu=Q)=\frac{(2\pi)^2}{\ln \frac{\Lambda_{\rm LP}^2}{Q^2}}
\ee
is a positive coupling whenever $Q^2<\Lambda_{\rm LP}^2$.

This seems to contradict our earlier conclusion that the infrared properties of the theory must be trivial, so let's find out how this happened. First, note that in the continuum limit
\be
\label{lambda0run}
\lim_{\Lambda_{\rm UV}\rightarrow \infty}\lambda_0\rightarrow 0^-\,,
\ee
implying that the bare-coupling constant should approach zero in the continuum limit from below. Since we know that the $\beta$-function of the theory (\ref{beta}) is always positive, and since the bare coupling is defined as the running coupling evaluated at the cut-off scale $\lambda_0=\lambda_R(\Lambda_{\rm UV})$, this would suggest that the coupling in the infrared has to be smaller than $\lambda_0$.

The standard interpretation, and the one that lead to our earlier quantum triviality result (\ref{qtriv}) is that the only way a coupling can grow from the infrared to the UV and reach zero at the end is that the whole function is zero. 

But that's not what happens here.

The actual running coupling $\lambda_R(\mu)$ in the continuum limit (\ref{contrun}) is shown in Fig.~\ref{fig1}. It is increasing in the whole domain $\mu \in [0,\infty)$, in agreement with the positive $\beta$-function (\ref{beta}), but it is non-analytic at the momentum scale $\mu=\Lambda_{\rm LP}$ where it possesses a Landau pole \cite{Romatschke:2022llf}, approaching zero from below in the UV, in agreement with Ref.~\cite{Linde:1976qh}.

Indeed, one of the assumptions that go into the all the available quantum triviality proofs e.g. those in Ref.~\cite{Aizenman:2019yuo} is that the bare coupling constant is positive, $\lambda_0>0$. This means that if one is willing to accept a negative bare coupling constant in the UV, quantum triviality can be avoided.

However, there is a price to pay.

The interacting theory has two unusual features: first, there is a Landau pole where the coupling constant $\lambda_R$ diverges located at $\mu=\Lambda_{\rm LP}$, and second, the coupling constant is negative above the Landau pole $\lambda_R<0$, suggesting a potential for the theory that is unbounded from below. I cannot resist pointing out that these unusual features are shared with the non-relativistic field theory of Fermions in a magnetic field interacting via s-wave scattering, where the scattering length takes the role of the coupling constant, cf. Fig. 2 in Ref.~\cite{gurarie2007resonantly}.

In the following, I will perform calculations for interacting N-component scalar theory in the continuum limit, in the presence of the Landau pole and the unbounded potential, showing that despite the unusual features, the properties of the theory are well-defined and reasonable.

\section{Scattering in Continuum $\phi^4$ theory}

\begin{figure}[t]
  \includegraphics[width=.95 \linewidth]{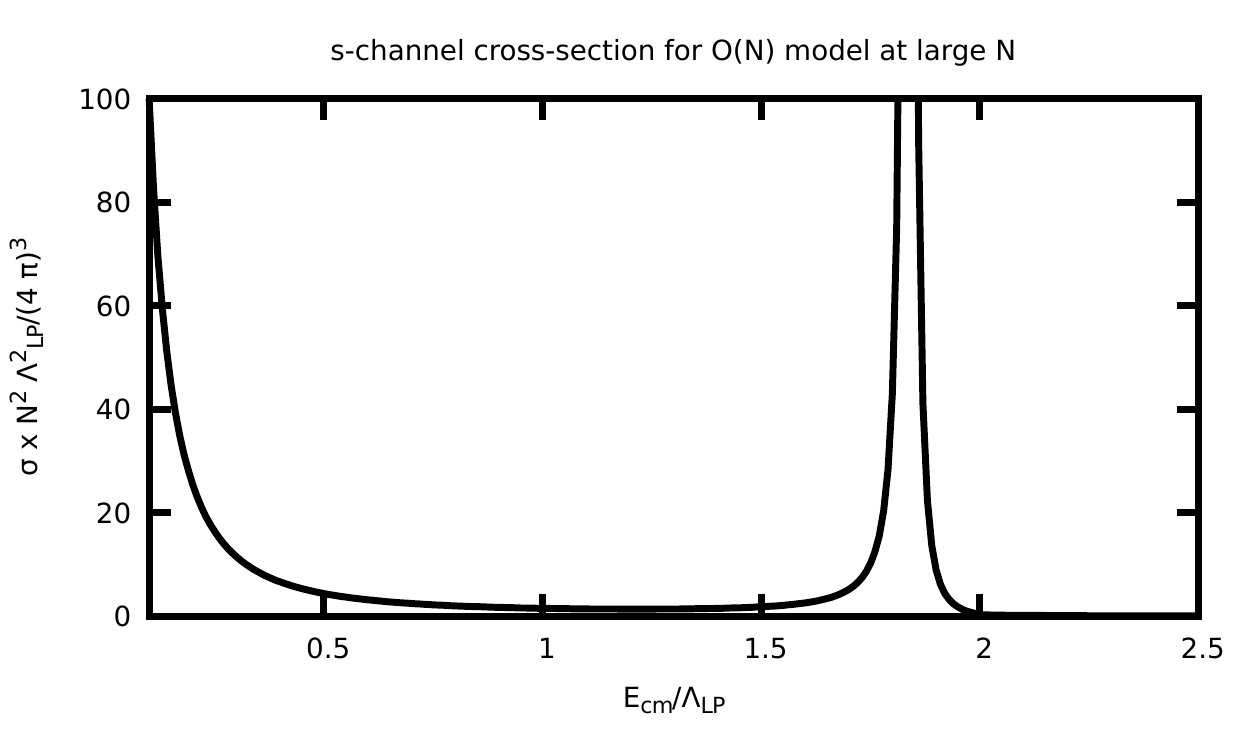}
  \caption{\label{fig2} Cross section in the s-channel for scattering in the continuum limit of large N scalar $(\vec{\phi}^2)^2$ theory. Note the presence of a bound state at $E\simeq 1.84 \Lambda_{\rm LP}$. See text for details.}
  \end{figure}

One can test if continuum $\phi^4$ theory is trivial by checking that the connected four-point function is Gaussian. For this purpose, let me consider the connected and amputated 4-point function
\be
\Gamma_4=-\langle \phi_a(x_1) \phi_b(x_2) \phi_c(x_3)\phi_d(x_4)\rangle_{\rm conn., amp.}\,,
\ee
where $\langle \cdot \rangle$ denotes the expectation value in the path integral representation and $a,b,c,d$ index components of the field $\phi$. Using the auxiliary field formulation from above implies that this expectation value is ${\cal O}(N^{-1})$ in the large N limit. Since this is NLO in the large N limit, this implies that the leading large N action is insufficient for performing this calculation correctly, and the NLO large N contribution in the action needs to be included. Fortunately, the technology to include the NLO term in the effective action is well developed, and in this case only requires the relatively simple resummation level R2 \footnote{For definitions and applications of R-level resummation schemes, see e.g. \cite{Romatschke:2019rjk,Romatschke:2019wxc,Romatschke:2021imm,Weiner:2022kgx}.}. R2 is defined by adding and subtracting a term
\be
\frac{1}{2} \int dx dy \zeta(x) \Pi(x-y) \zeta(y)\,,
\ee
to the action (\ref{action}). Working to leading order in large N then leads to the self-energy $\Pi(x)$ given by
\be
\label{pif}
\Pi(x)=\frac{N}{2}G^2(x)\,,
\ee
with $G(x)$ the $\vec{\phi}$ propagator which in Fourier space is given by $G(k)=\frac{1}{k^2+m_0^2+z^*}$. The connected amputated four-point function is given by a sum of $s,t$ and $u$ channel contributions, all of which have the same form. For instance, the s-channel contribution is
\be
\Gamma_4^{\rm (s)}(p)=\frac{1}{\frac{N}{8\lambda_0}+\Pi(p)}\,,
\ee
where $p$ is the sum of the incoming momenta.

Again, it should be stressed that our expansion parameter is $\frac{1}{N}\ll 1$, and not a small coupling expansion. If doing a naive small-coupling expansion of $\Gamma_4$, the 4-point function would behave as $\Gamma_4\propto \frac{\lambda_0}{N}$, which vanishes in the continuum because of (\ref{lambda0run}), suggesting again quantum triviality. However, this is an artifact of using perturbation theory, because the non-perturbative result behaves quite differently.

Evaluating the self-energy (\ref{pif}) as in Ref.~\cite{Romatschke:2022jqg}, one finds in the continuum limit
\be
\label{g4s}
\Gamma_4^{(s)}(p)=\frac{(2 \pi)^2}{\frac{N}{8}\left(\frac{(2 \pi)^2}{\lambda_R(\mu)}+\ln\frac{\mu^2 e}{M^2}-2 \sqrt{\frac{p^2+4 M^2}{p^2}}{\rm atanh} \sqrt{\frac{p^2}{p^2+4 M^2}}\right)}\,,
\ee
where $M^2=m_0^2+z^*=\mu^2 e^{\frac{(2\pi)^2}{\lambda_R(\mu)}}=\Lambda_{LP}^2$ at the non-perturbative saddle when using (\ref{contrun}).

From (\ref{g4s}), it becomes clear that $\Gamma_4$ is non-vanishing in the continuum limit. This is also born out by the scattering cross-section, which in the s-channel becomes
\be
\sigma^{(s)}(E)=\frac{(4 \pi)^3}{N^2 E^2 \left|1-2 \sqrt{1-\frac{4 M^2}{E^2+i 0^+}}{\rm atanh} \frac{1}{\sqrt{1-\frac{4 M^2}{E^2+i 0^+}}}\right|^2}\,.
\ee
A plot of the s-channel cross-section is shown in Fig.~\ref{fig2}. From this plot,one can clearly see the presence of a bound state with a mass of approximately $1.84$ M, which is another indication of non-trivial behavior of scalar $\phi^4$ theory in the large N limit\footnote{I thank Paolo Cea for pointing out that my auxiliary field $\zeta$ and its bound state are unrelated to the BEC and second Higgs reported in Ref.~\cite{Cea:2020lea}.}. (Note that my result is in full agreement with an earlier calculation \cite{Abbott:1975bn}.) I expect that $\frac{1}{N}$ corrections at R3 resummation level will turn the stable bound state into a resonance, but calculating the width along the lines of \cite{Romatschke:2021imm} is beyond the scope of this work.

Other observables of this theory are similarly well-defined despite the presence of the Landau pole. For instance, putting the theory at finite temperature one finds that the low-temperature phase is separated from a high-temperature phase by a second-order phase transition \cite{Romatschke:2022jqg}, and the thermodynamics are qualitatively similar to that of real-world QCD \cite{Romatschke:2022llf}.

\section{$\phi^4$ theory with $\lambda_0<0$ on the lattice}

Instead of multi-component scalar fields, let me now consider the case of a single scalar field with quartic interaction, with partition function $Z$ given by (\ref{Z1}) with $N=1$. Since the previous arguments suggest that the bare coupling parameter $\lambda_0$ in the continuum limit should be negative, the path integral is ill defined when the integration domain for the fields $\phi(x)$ are real. However, borrowing from technology developed in the context of ${\cal PT}$-symmetric quantum mechanics, one can contour-deform the integration domain to be complex. For instance, one can contour-deform to the 'cone' contour \cite{Lawrence:2023woz} such that each field $\phi(x)$ is parametrized by
\be
\label{cone}
\phi(x)=s(x)\left(e^{\frac{i \pi}{4}}\Theta(-s(x))+e^{\frac{-i \pi}{4}}\Theta(s(x))\right)\,,
\ee
where $s(x)$ is taken to be real. (Note that smoother contours for the same purpose exist, cf.~\cite{Jones:2006qs}). It should be pointed out that in terms of the real fields $s(x)$, the action is non-polynomial because of the distributional property in (\ref{cone}), thereby evading the mathematical proofs for quantum triviality and $N=1$.

\begin{figure}[t]
  \includegraphics[width=.95 \linewidth]{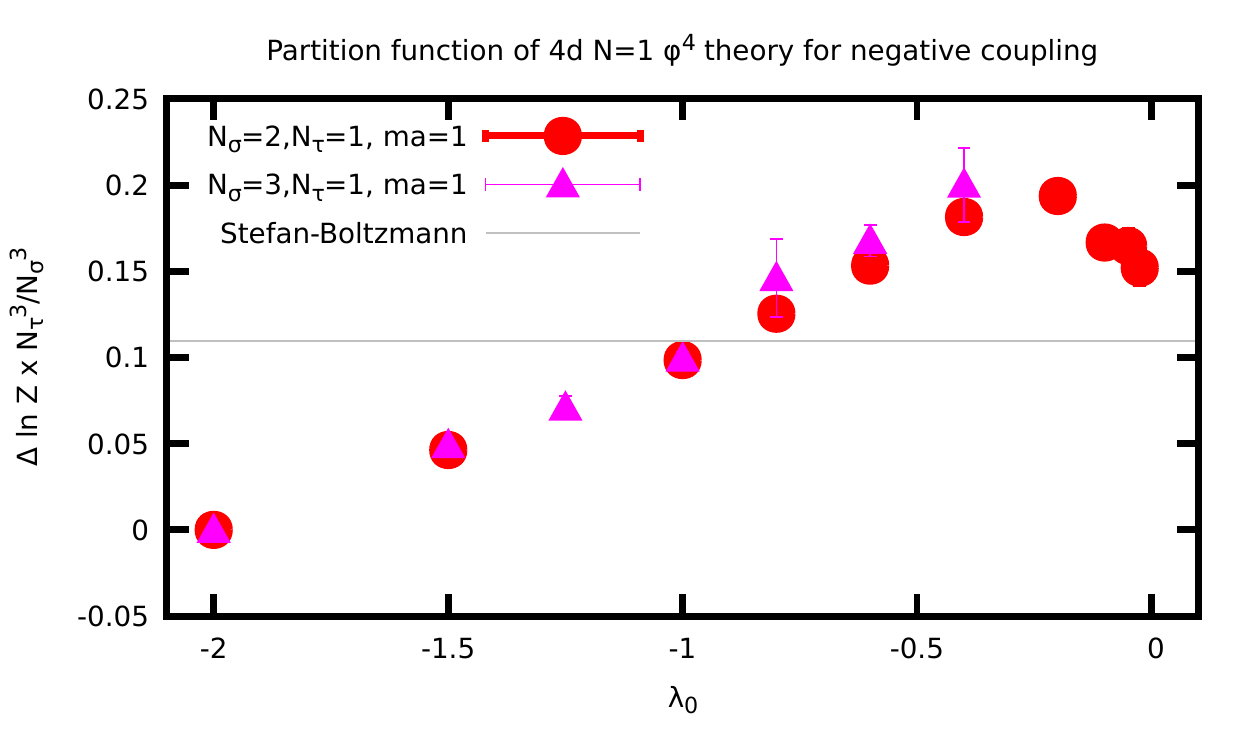}
  \caption{\label{fig3} Partition function for scalar $\phi^4$ theory with negative bare coupling on small lattices. Shown are $\Delta \ln Z(\lambda_0)=\ln \frac{Z(\lambda_0)}{Z(\lambda_0=-2)}$ and the Stefan-Boltzmann constant for a single bosonic degree of freedom. Since the subtraction point $\lambda_0=-2$ is chosen ad-hoc, the apparent similarity between simulations and the Stefan-Boltzmann constant is entirely accidental. See text for details.}
  \end{figure}

With this complex reparametrization, one can discretize the theory on a hypercubic $N_\sigma^3\times N_\tau$ lattice where $N_{\sigma},N_\tau$ represent the points in the spatial and temporal directions, respectively. In terms of the real-valued discretized fields $s_i$, the lattice action $S_L$ is complex, but symmetry guarantees that the partition function $Z\propto \int \prod_{i=1}^{\rm sites} ds_i e^{-S_L}$ is real-valued. In addition, on the complex cone the lattice action contains a term $S_L\propto - \lambda_0 \sum_{i=1}^{\rm sites} s_i^4$ that guarantees that the discretized path integral converges for $\lambda_0<0$. Unfortunately, the complex-valued action renders importance sampling of the path integral unviable, but for low enough dimensions the partition function is amenable to direct integration \cite{Lepage:2020tgj}. Choosing a bare mass parameter of $m_0=1$ in lattice unites for simplicity, a plot of $\ln Z$ is shown in Fig.~\ref{fig3} for $N_\sigma=2,3$ and $N_\tau=1$.

The observable shown in this figures corresponds to the dimensionless pressure of the theory $\frac{P(T)-P(T_0)}{T^4}=\frac{\Delta \ln Z}{V T^3}$, where on the lattice with lattice spacing $a$ the temperature is given by $T=\frac{1}{N_\tau a}$ and $T_0$ is some reference temperature, which I arbitrarily choose here as corresponding to $\lambda_0=-2$. Identifying $a=\frac{1}{\Lambda_{\rm UV}}$, one expects from (\ref{lambda0}) that $\lambda_0\rightarrow 0^-$ corresponds to the high temperature limit of the theory. From the large N results from \cite{Romatschke:2022jqg}, one expects $\frac{\Delta P}{T^4}$ to first overshoot the Stefan-Boltzmann constant, and then decrease with decreasing temperature, which is indeed what is found in Fig.~\ref{fig3}. Unfortunately, the small lattice sizes prohibit the investigation of a possible second-order phase transition that characterizes the large N result in 
\cite{Romatschke:2022jqg}. Nevertheless, the results show that it is in principle possible to study $\phi^4$ theory with negative coupling on the lattice and obtain non-perturbative information in this fashion.

\section{Conclusions}

While the existence of scalar field theory was seriously debated in the 1970s\cite{Schnitzer:1974ji,Coleman:1974jh,Abbott:1975bn,Linde:1976qh,Linde:1977mm}, the dominating point of view in the past few decades seems to have become that scalar field theory in four dimensions in the continuum is trivial. In this work, I boldly challenged this consensus point of view.

I pointed out that mathematical proofs on the triviality of scalar field theory in the continuum do not cover certain situations of interest, notably multi-component scalar fields and non-polynomial action.

In this work, I have studied N-component scalar field theory in the continuum and large N limits, finding a non-trivial theory with interesting properties, such as a resonance with a mass of $\sim 1.84$ times the scalar mass. Since mathematical proofs of triviality are limited to $N\leq 2$, my results do not contradict any rigorously proven theorems. Amusingly, this means that the Standard Model Higgs field, which is N=4, would not be covered by triviality proofs, but possibly by the large N result I presented in this work.

If results apply to the Standard Model Higgs,  I predict a (possibly broad) Higgs-pair resonance with a mass of about $M\simeq 1.84 m_H\simeq 230$ GeV to be present in the Standard Model.

Setting aside the case of multi-component fields, I have also investigated the case of regular (one-component) scalar $\phi^4$ field theory. Since this theory cannot be accessed in the large N framework, I have used direct numerical simulations of $\phi^4$ theory with negative bare coupling on the lattice. Limited by numerics, my results for N=1 scalar field theory are too weak to make claims about triviality of single component $\phi^4$ theory, but using contour deformation I found that the resulting non-polynomial field theory can be studied numerically, with encouraging results.

The implications of my result are possibly very broad, as they directly impact the reigning EFT paradigm of quantum field theory. Also, if scalar $\phi^4$ theory is nontrivial in the continuum, so could be many more quantum field theories, which in turn suggest that there is a whole new class of quantum field theories that can be studied. At the technical level, there are several extensions of this work that should be explored in the future, such as calculating $\frac{1}{N}$ corrections to the scalar bound state result, and numerical experiments on much larger lattices for the single-component scalar field.

A top priority, however, is to study how the results presented here affect and possibly modify the electroweak sector of the Standard Model of physics.

\section*{Acknowledgments}

This work was supported by the Department of Energy, DOE award No DE-SC0017905.  I would like to thank Scott Lawrence for many illuminating discussions, Slava Rychkov and Poul Damgaard for encouraging comments, Paolo Cea for clarifications and Andrei Linde and Howard Schnitzer for pointing me to some of the pioneering results on this topic in the 1970s.

\bibliography{PT}

\begin{thebibliography}{35}%
\makeatletter
\providecommand \@ifxundefined [1]{%
 \@ifx{#1\undefined}
}%
\providecommand \@ifnum [1]{%
 \ifnum #1\expandafter \@firstoftwo
 \else \expandafter \@secondoftwo
 \fi
}%
\providecommand \@ifx [1]{%
 \ifx #1\expandafter \@firstoftwo
 \else \expandafter \@secondoftwo
 \fi
}%
\providecommand \natexlab [1]{#1}%
\providecommand \enquote  [1]{``#1''}%
\providecommand \bibnamefont  [1]{#1}%
\providecommand \bibfnamefont [1]{#1}%
\providecommand \citenamefont [1]{#1}%
\providecommand \href@noop [0]{\@secondoftwo}%
\providecommand \href [0]{\begingroup \@sanitize@url \@href}%
\providecommand \@href[1]{\@@startlink{#1}\@@href}%
\providecommand \@@href[1]{\endgroup#1\@@endlink}%
\providecommand \@sanitize@url [0]{\catcode `\\12\catcode `\$12\catcode
  `\&12\catcode `\#12\catcode `\^12\catcode `\_12\catcode `\%12\relax}%
\providecommand \@@startlink[1]{}%
\providecommand \@@endlink[0]{}%
\providecommand \url  [0]{\begingroup\@sanitize@url \@url }%
\providecommand \@url [1]{\endgroup\@href {#1}{\urlprefix }}%
\providecommand \urlprefix  [0]{URL }%
\providecommand \Eprint [0]{\href }%
\providecommand \doibase [0]{http://dx.doi.org/}%
\providecommand \selectlanguage [0]{\@gobble}%
\providecommand \bibinfo  [0]{\@secondoftwo}%
\providecommand \bibfield  [0]{\@secondoftwo}%
\providecommand \translation [1]{[#1]}%
\providecommand \BibitemOpen [0]{}%
\providecommand \bibitemStop [0]{}%
\providecommand \bibitemNoStop [0]{.\EOS\space}%
\providecommand \EOS [0]{\spacefactor3000\relax}%
\providecommand \BibitemShut  [1]{\csname bibitem#1\endcsname}%
\let\auto@bib@innerbib\@empty
\bibitem [{\citenamefont {Wilson}\ and\ \citenamefont
  {Kogut}(1974)}]{Wilson:1973jj}%
  \BibitemOpen
  \bibfield  {author} {\bibinfo {author} {\bibfnamefont {K.~G.}\ \bibnamefont
  {Wilson}}\ and\ \bibinfo {author} {\bibfnamefont {John~B.}\ \bibnamefont
  {Kogut}},\ }\bibfield  {title} {\enquote {\bibinfo {title} {{The
  Renormalization group and the epsilon expansion}},}\ }\href {\doibase
  10.1016/0370-1573(74)90023-4} {\bibfield  {journal} {\bibinfo  {journal}
  {Phys. Rept.}\ }\textbf {\bibinfo {volume} {12}},\ \bibinfo {pages} {75--199}
  (\bibinfo {year} {1974})}\BibitemShut {NoStop}%
\bibitem [{\citenamefont {Frohlich}(1982)}]{Frohlich:1982tw}%
  \BibitemOpen
  \bibfield  {author} {\bibinfo {author} {\bibfnamefont {J.}~\bibnamefont
  {Frohlich}},\ }\bibfield  {title} {\enquote {\bibinfo {title} {{On the
  Triviality of Lambda (phi**4) in D-Dimensions Theories and the Approach to
  the Critical Point in D \ensuremath{>}= Four-Dimensions}},}\ }\href {\doibase
  10.1016/0550-3213(82)90088-8} {\bibfield  {journal} {\bibinfo  {journal}
  {Nucl. Phys. B}\ }\textbf {\bibinfo {volume} {200}},\ \bibinfo {pages}
  {281--296} (\bibinfo {year} {1982})}\BibitemShut {NoStop}%
\bibitem [{\citenamefont {Aizenman}\ and\ \citenamefont
  {Duminil-Copin}(2021)}]{Aizenman:2019yuo}%
  \BibitemOpen
  \bibfield  {author} {\bibinfo {author} {\bibfnamefont {Michael}\ \bibnamefont
  {Aizenman}}\ and\ \bibinfo {author} {\bibfnamefont {Hugo}\ \bibnamefont
  {Duminil-Copin}},\ }\bibfield  {title} {\enquote {\bibinfo {title} {{Marginal
  triviality of the scaling limits of critical 4D Ising and $\phi_4^4$
  models}},}\ }\href {\doibase 10.4007/annals.2021.194.1.3} {\bibfield
  {journal} {\bibinfo  {journal} {Annals Math.}\ }\textbf {\bibinfo {volume}
  {194}},\ \bibinfo {pages} {163} (\bibinfo {year} {2021})},\ \Eprint
  {http://arxiv.org/abs/1912.07973} {arXiv:1912.07973 [math-ph]} \BibitemShut
  {NoStop}%
\bibitem [{\citenamefont {Bender}\ and\ \citenamefont
  {Boettcher}(1998)}]{Bender:1998ke}%
  \BibitemOpen
  \bibfield  {author} {\bibinfo {author} {\bibfnamefont {Carl~M.}\ \bibnamefont
  {Bender}}\ and\ \bibinfo {author} {\bibfnamefont {Stefan}\ \bibnamefont
  {Boettcher}},\ }\bibfield  {title} {\enquote {\bibinfo {title} {{Real spectra
  in non-Hermitian Hamiltonians having PT symmetry}},}\ }\href {\doibase
  10.1103/PhysRevLett.80.5243} {\bibfield  {journal} {\bibinfo  {journal}
  {Phys. Rev. Lett.}\ }\textbf {\bibinfo {volume} {80}},\ \bibinfo {pages}
  {5243--5246} (\bibinfo {year} {1998})},\ \Eprint
  {http://arxiv.org/abs/physics/9712001} {arXiv:physics/9712001} \BibitemShut
  {NoStop}%
\bibitem [{\citenamefont {Dorey}\ \emph {et~al.}(2001)\citenamefont {Dorey},
  \citenamefont {Dunning},\ and\ \citenamefont {Tateo}}]{Dorey:2001uw}%
  \BibitemOpen
  \bibfield  {author} {\bibinfo {author} {\bibfnamefont {Patrick}\ \bibnamefont
  {Dorey}}, \bibinfo {author} {\bibfnamefont {Clare}\ \bibnamefont {Dunning}},
  \ and\ \bibinfo {author} {\bibfnamefont {Roberto}\ \bibnamefont {Tateo}},\
  }\bibfield  {title} {\enquote {\bibinfo {title} {{Spectral equivalences,
  Bethe Ansatz equations, and reality properties in PT-symmetric quantum
  mechanics}},}\ }\href {\doibase 10.1088/0305-4470/34/28/305} {\bibfield
  {journal} {\bibinfo  {journal} {J. Phys. A}\ }\textbf {\bibinfo {volume}
  {34}},\ \bibinfo {pages} {5679--5704} (\bibinfo {year} {2001})},\ \Eprint
  {http://arxiv.org/abs/hep-th/0103051} {arXiv:hep-th/0103051} \BibitemShut
  {NoStop}%
\bibitem [{\citenamefont {Bender}\ \emph {et~al.}(2021)\citenamefont {Bender},
  \citenamefont {Felski}, \citenamefont {Klevansky},\ and\ \citenamefont
  {Sarkar}}]{Bender:2021fxa}%
  \BibitemOpen
  \bibfield  {author} {\bibinfo {author} {\bibfnamefont {Carl~M.}\ \bibnamefont
  {Bender}}, \bibinfo {author} {\bibfnamefont {Alexander}\ \bibnamefont
  {Felski}}, \bibinfo {author} {\bibfnamefont {S.~P.}\ \bibnamefont
  {Klevansky}}, \ and\ \bibinfo {author} {\bibfnamefont {Sarben}\ \bibnamefont
  {Sarkar}},\ }\bibfield  {title} {\enquote {\bibinfo {title} {{$PT$ Symmetry
  and Renormalisation in Quantum Field Theory}},}\ }\href {\doibase
  10.1088/1742-6596/2038/1/012004} {\bibfield  {journal} {\bibinfo  {journal}
  {J. Phys. Conf. Ser.}\ }\textbf {\bibinfo {volume} {2038}},\ \bibinfo {pages}
  {012004} (\bibinfo {year} {2021})},\ \Eprint
  {http://arxiv.org/abs/2103.14864} {arXiv:2103.14864 [hep-th]} \BibitemShut
  {NoStop}%
\bibitem [{\citenamefont {Mavromatos}\ \emph {et~al.}(2022)\citenamefont
  {Mavromatos}, \citenamefont {Sarkar},\ and\ \citenamefont
  {Soto}}]{Mavromatos:2021hpe}%
  \BibitemOpen
  \bibfield  {author} {\bibinfo {author} {\bibfnamefont {N.~E.}\ \bibnamefont
  {Mavromatos}}, \bibinfo {author} {\bibfnamefont {Sarben}\ \bibnamefont
  {Sarkar}}, \ and\ \bibinfo {author} {\bibfnamefont {A.}~\bibnamefont
  {Soto}},\ }\bibfield  {title} {\enquote {\bibinfo {title} {{PT symmetric
  fermionic field theories with axions: Renormalization and dynamical mass
  generation}},}\ }\href {\doibase 10.1103/PhysRevD.106.015009} {\bibfield
  {journal} {\bibinfo  {journal} {Phys. Rev. D}\ }\textbf {\bibinfo {volume}
  {106}},\ \bibinfo {pages} {015009} (\bibinfo {year} {2022})},\ \Eprint
  {http://arxiv.org/abs/2111.05131} {arXiv:2111.05131 [hep-th]} \BibitemShut
  {NoStop}%
\bibitem [{\citenamefont {Grunwald}\ \emph {et~al.}(2022)\citenamefont
  {Grunwald}, \citenamefont {Meden},\ and\ \citenamefont
  {Kennes}}]{Grunwald:2022kts}%
  \BibitemOpen
  \bibfield  {author} {\bibinfo {author} {\bibfnamefont {Lukas}\ \bibnamefont
  {Grunwald}}, \bibinfo {author} {\bibfnamefont {Volker}\ \bibnamefont
  {Meden}}, \ and\ \bibinfo {author} {\bibfnamefont {Dante~M.}\ \bibnamefont
  {Kennes}},\ }\bibfield  {title} {\enquote {\bibinfo {title} {{Functional
  renormalization group for non-Hermitian and $\mathcal{PT}$-symmetric
  systems}},}\ }\href {\doibase 10.21468/SciPostPhys.12.5.179} {\bibfield
  {journal} {\bibinfo  {journal} {SciPost Phys.}\ }\textbf {\bibinfo {volume}
  {12}},\ \bibinfo {pages} {179} (\bibinfo {year} {2022})},\ \Eprint
  {http://arxiv.org/abs/2203.08108} {arXiv:2203.08108 [cond-mat.str-el]}
  \BibitemShut {NoStop}%
\bibitem [{\citenamefont {Ai}\ \emph {et~al.}(2022)\citenamefont {Ai},
  \citenamefont {Bender},\ and\ \citenamefont {Sarkar}}]{Ai:2022csx}%
  \BibitemOpen
  \bibfield  {author} {\bibinfo {author} {\bibfnamefont {Wen-Yuan}\
  \bibnamefont {Ai}}, \bibinfo {author} {\bibfnamefont {Carl~M.}\ \bibnamefont
  {Bender}}, \ and\ \bibinfo {author} {\bibfnamefont {Sarben}\ \bibnamefont
  {Sarkar}},\ }\bibfield  {title} {\enquote {\bibinfo {title} {{PT-symmetric
  -g\ensuremath{\varphi}4 theory}},}\ }\href {\doibase
  10.1103/PhysRevD.106.125016} {\bibfield  {journal} {\bibinfo  {journal}
  {Phys. Rev. D}\ }\textbf {\bibinfo {volume} {106}},\ \bibinfo {pages}
  {125016} (\bibinfo {year} {2022})},\ \Eprint
  {http://arxiv.org/abs/2209.07897} {arXiv:2209.07897 [hep-th]} \BibitemShut
  {NoStop}%
\bibitem [{\citenamefont
  {Romatschke}(2022{\natexlab{a}})}]{Romatschke:2022jqg}%
  \BibitemOpen
  \bibfield  {author} {\bibinfo {author} {\bibfnamefont {Paul}\ \bibnamefont
  {Romatschke}},\ }\bibfield  {title} {\enquote {\bibinfo {title} {{A solvable
  quantum field theory with asymptotic freedom in 3+1 dimensions}},}\
  }\href@noop {} {\  (\bibinfo {year} {2022}{\natexlab{a}})},\ \Eprint
  {http://arxiv.org/abs/2211.15683} {arXiv:2211.15683 [hep-th]} \BibitemShut
  {NoStop}%
\bibitem [{\citenamefont {Grable}\ and\ \citenamefont
  {Weiner}(2023)}]{Grable:2023paf}%
  \BibitemOpen
  \bibfield  {author} {\bibinfo {author} {\bibfnamefont {Seth}\ \bibnamefont
  {Grable}}\ and\ \bibinfo {author} {\bibfnamefont {Max}\ \bibnamefont
  {Weiner}},\ }\bibfield  {title} {\enquote {\bibinfo {title} {{A Fully
  Solvable Model of Fermionic Interaction in $3+1d$}},}\ }\href@noop {} {\
  (\bibinfo {year} {2023})},\ \Eprint {http://arxiv.org/abs/2302.08603}
  {arXiv:2302.08603 [hep-th]} \BibitemShut {NoStop}%
\bibitem [{\citenamefont {Lawrence}\ \emph {et~al.}(2023)\citenamefont
  {Lawrence}, \citenamefont {Weller}, \citenamefont {Peterson},\ and\
  \citenamefont {Romatschke}}]{Lawrence:2023woz}%
  \BibitemOpen
  \bibfield  {author} {\bibinfo {author} {\bibfnamefont {Scott}\ \bibnamefont
  {Lawrence}}, \bibinfo {author} {\bibfnamefont {Ryan}\ \bibnamefont {Weller}},
  \bibinfo {author} {\bibfnamefont {Christian}\ \bibnamefont {Peterson}}, \
  and\ \bibinfo {author} {\bibfnamefont {Paul}\ \bibnamefont {Romatschke}},\
  }\bibfield  {title} {\enquote {\bibinfo {title} {{Instantons, analytic
  continuation, and $\mathcal{PT}$-symmetric field theory}},}\ }\href@noop {}
  {\  (\bibinfo {year} {2023})},\ \Eprint {http://arxiv.org/abs/2303.01470}
  {arXiv:2303.01470 [hep-th]} \BibitemShut {NoStop}%
\bibitem [{\citenamefont
  {Romatschke}(2022{\natexlab{b}})}]{Romatschke:2022llf}%
  \BibitemOpen
  \bibfield  {author} {\bibinfo {author} {\bibfnamefont {Paul}\ \bibnamefont
  {Romatschke}},\ }\bibfield  {title} {\enquote {\bibinfo {title} {{Life at the
  Landau pole}},}\ }\href@noop {} {\  (\bibinfo {year} {2022}{\natexlab{b}})},\
  \Eprint {http://arxiv.org/abs/2212.03254} {arXiv:2212.03254 [hep-th]}
  \BibitemShut {NoStop}%
\bibitem [{Note1()}]{Note1}%
  \BibitemOpen
  \bibinfo {note} {For a very incomplete list of examples in various field
  theories, cf. Refs.~\cite
  {Parisi:1975im,Itzhaki:1998dd,Klebanov:2002ja,Romatschke:2019ybu,Romatschke:2021imm}.}\BibitemShut
  {Stop}%
\bibitem [{\citenamefont {Symanzik}(1973)}]{Symanzik:1973hx}%
  \BibitemOpen
  \bibfield  {author} {\bibinfo {author} {\bibfnamefont {K.}~\bibnamefont
  {Symanzik}},\ }\bibfield  {title} {\enquote {\bibinfo {title} {{A field
  theory with computable large-momenta behavior}},}\ }\href {\doibase
  10.1007/BF02788323} {\bibfield  {journal} {\bibinfo  {journal} {Lett. Nuovo
  Cim.}\ }\textbf {\bibinfo {volume} {6S2}},\ \bibinfo {pages} {77--80}
  (\bibinfo {year} {1973})}\BibitemShut {NoStop}%
\bibitem [{\citenamefont {Moshe}\ and\ \citenamefont
  {Zinn-Justin}(2003)}]{Moshe:2003xn}%
  \BibitemOpen
  \bibfield  {author} {\bibinfo {author} {\bibfnamefont {Moshe}\ \bibnamefont
  {Moshe}}\ and\ \bibinfo {author} {\bibfnamefont {Jean}\ \bibnamefont
  {Zinn-Justin}},\ }\bibfield  {title} {\enquote {\bibinfo {title} {{Quantum
  field theory in the large N limit: A Review}},}\ }\href {\doibase
  10.1016/S0370-1573(03)00263-1} {\bibfield  {journal} {\bibinfo  {journal}
  {Phys. Rept.}\ }\textbf {\bibinfo {volume} {385}},\ \bibinfo {pages}
  {69--228} (\bibinfo {year} {2003})},\ \Eprint
  {http://arxiv.org/abs/hep-th/0306133} {arXiv:hep-th/0306133} \BibitemShut
  {NoStop}%
\bibitem [{\citenamefont
  {Romatschke}(2019{\natexlab{a}})}]{Romatschke:2019rjk}%
  \BibitemOpen
  \bibfield  {author} {\bibinfo {author} {\bibfnamefont {Paul}\ \bibnamefont
  {Romatschke}},\ }\bibfield  {title} {\enquote {\bibinfo {title} {{Simple
  non-perturbative resummation schemes beyond mean-field: case study for scalar
  $\phi^4$ theory in 1+1 dimensions}},}\ }\href {\doibase
  10.1007/JHEP03(2019)149} {\bibfield  {journal} {\bibinfo  {journal} {JHEP}\
  }\textbf {\bibinfo {volume} {03}},\ \bibinfo {pages} {149} (\bibinfo {year}
  {2019}{\natexlab{a}})},\ \Eprint {http://arxiv.org/abs/1901.05483}
  {arXiv:1901.05483 [hep-th]} \BibitemShut {NoStop}%
\bibitem [{\citenamefont {Linde}(1977{\natexlab{a}})}]{Linde:1976qh}%
  \BibitemOpen
  \bibfield  {author} {\bibinfo {author} {\bibfnamefont {Andrei~D.}\
  \bibnamefont {Linde}},\ }\bibfield  {title} {\enquote {\bibinfo {title}
  {{1/n-Expansion, Vacuum Stability and Quark Confinement}},}\ }\href {\doibase
  10.1016/0550-3213(77)90112-2} {\bibfield  {journal} {\bibinfo  {journal}
  {Nucl. Phys. B}\ }\textbf {\bibinfo {volume} {125}},\ \bibinfo {pages}
  {369--380} (\bibinfo {year} {1977}{\natexlab{a}})}\BibitemShut {NoStop}%
\bibitem [{\citenamefont {Gurarie}\ and\ \citenamefont
  {Radzihovsky}(2007)}]{gurarie2007resonantly}%
  \BibitemOpen
  \bibfield  {author} {\bibinfo {author} {\bibfnamefont {V}~\bibnamefont
  {Gurarie}}\ and\ \bibinfo {author} {\bibfnamefont {L}~\bibnamefont
  {Radzihovsky}},\ }\bibfield  {title} {\enquote {\bibinfo {title} {Resonantly
  paired fermionic superfluids},}\ }\href@noop {} {\bibfield  {journal}
  {\bibinfo  {journal} {Annals of Physics}\ }\textbf {\bibinfo {volume}
  {322}},\ \bibinfo {pages} {2--119} (\bibinfo {year} {2007})}\BibitemShut
  {NoStop}%
\bibitem [{Note2()}]{Note2}%
  \BibitemOpen
  \bibinfo {note} {For definitions and applications of R-level resummation
  schemes, see e.g. \cite
  {Romatschke:2019rjk,Romatschke:2019wxc,Romatschke:2021imm,Weiner:2022kgx}.}\BibitemShut
  {Stop}%
\bibitem [{Note3()}]{Note3}%
  \BibitemOpen
  \bibinfo {note} {I thank Paolo Cea for pointing out that my auxiliary field
  $\zeta $ and its bound state are unrelated to the BEC and second Higgs
  reported in Ref.~\cite {Cea:2020lea}.}\BibitemShut {Stop}%
\bibitem [{\citenamefont {Abbott}\ \emph {et~al.}(1976)\citenamefont {Abbott},
  \citenamefont {Kang},\ and\ \citenamefont {Schnitzer}}]{Abbott:1975bn}%
  \BibitemOpen
  \bibfield  {author} {\bibinfo {author} {\bibfnamefont {L.~F.}\ \bibnamefont
  {Abbott}}, \bibinfo {author} {\bibfnamefont {J.~S.}\ \bibnamefont {Kang}}, \
  and\ \bibinfo {author} {\bibfnamefont {Howard~J.}\ \bibnamefont
  {Schnitzer}},\ }\bibfield  {title} {\enquote {\bibinfo {title} {{Bound
  States, Tachyons, and Restoration of Symmetry in the 1/N Expansion}},}\
  }\href {\doibase 10.1103/PhysRevD.13.2212} {\bibfield  {journal} {\bibinfo
  {journal} {Phys. Rev. D}\ }\textbf {\bibinfo {volume} {13}},\ \bibinfo
  {pages} {2212} (\bibinfo {year} {1976})}\BibitemShut {NoStop}%
\bibitem [{\citenamefont {Romatschke}(2021)}]{Romatschke:2021imm}%
  \BibitemOpen
  \bibfield  {author} {\bibinfo {author} {\bibfnamefont {Paul}\ \bibnamefont
  {Romatschke}},\ }\bibfield  {title} {\enquote {\bibinfo {title} {{Shear
  Viscosity at Infinite Coupling: A Field Theory Calculation}},}\ }\href
  {\doibase 10.1103/PhysRevLett.127.111603} {\bibfield  {journal} {\bibinfo
  {journal} {Phys. Rev. Lett.}\ }\textbf {\bibinfo {volume} {127}},\ \bibinfo
  {pages} {111603} (\bibinfo {year} {2021})},\ \Eprint
  {http://arxiv.org/abs/2104.06435} {arXiv:2104.06435 [hep-th]} \BibitemShut
  {NoStop}%
\bibitem [{\citenamefont {Jones}\ and\ \citenamefont
  {Mateo}(2006)}]{Jones:2006qs}%
  \BibitemOpen
  \bibfield  {author} {\bibinfo {author} {\bibfnamefont {H.~F.}\ \bibnamefont
  {Jones}}\ and\ \bibinfo {author} {\bibfnamefont {J.}~\bibnamefont {Mateo}},\
  }\bibfield  {title} {\enquote {\bibinfo {title} {{An Equivalent Hermitian
  Hamiltonian for the non-Hermitian -x**4 potential}},}\ }\href {\doibase
  10.1103/PhysRevD.73.085002} {\bibfield  {journal} {\bibinfo  {journal} {Phys.
  Rev. D}\ }\textbf {\bibinfo {volume} {73}},\ \bibinfo {pages} {085002}
  (\bibinfo {year} {2006})},\ \Eprint {http://arxiv.org/abs/quant-ph/0601188}
  {arXiv:quant-ph/0601188} \BibitemShut {NoStop}%
\bibitem [{\citenamefont {Lepage}(2021)}]{Lepage:2020tgj}%
  \BibitemOpen
  \bibfield  {author} {\bibinfo {author} {\bibfnamefont {G.~Peter}\
  \bibnamefont {Lepage}},\ }\bibfield  {title} {\enquote {\bibinfo {title}
  {{Adaptive multidimensional integration: VEGAS enhanced}},}\ }\href {\doibase
  10.1016/j.jcp.2021.110386} {\bibfield  {journal} {\bibinfo  {journal} {J.
  Comput. Phys.}\ }\textbf {\bibinfo {volume} {439}},\ \bibinfo {pages}
  {110386} (\bibinfo {year} {2021})},\ \Eprint
  {http://arxiv.org/abs/2009.05112} {arXiv:2009.05112 [physics.comp-ph]}
  \BibitemShut {NoStop}%
\bibitem [{\citenamefont {Schnitzer}(1974)}]{Schnitzer:1974ji}%
  \BibitemOpen
  \bibfield  {author} {\bibinfo {author} {\bibfnamefont {Howard~J.}\
  \bibnamefont {Schnitzer}},\ }\bibfield  {title} {\enquote {\bibinfo {title}
  {{Nonperturbative Effective Potential for Lambda phi**4 Theory in the Many
  Field Limit}},}\ }\href {\doibase 10.1103/PhysRevD.10.1800} {\bibfield
  {journal} {\bibinfo  {journal} {Phys. Rev. D}\ }\textbf {\bibinfo {volume}
  {10}},\ \bibinfo {pages} {1800} (\bibinfo {year} {1974})}\BibitemShut
  {NoStop}%
\bibitem [{\citenamefont {Coleman}\ \emph {et~al.}(1974)\citenamefont
  {Coleman}, \citenamefont {Jackiw},\ and\ \citenamefont
  {Politzer}}]{Coleman:1974jh}%
  \BibitemOpen
  \bibfield  {author} {\bibinfo {author} {\bibfnamefont {Sidney~R.}\
  \bibnamefont {Coleman}}, \bibinfo {author} {\bibfnamefont {R.}~\bibnamefont
  {Jackiw}}, \ and\ \bibinfo {author} {\bibfnamefont {H.~David}\ \bibnamefont
  {Politzer}},\ }\bibfield  {title} {\enquote {\bibinfo {title} {{Spontaneous
  Symmetry Breaking in the O(N) Model for Large N*}},}\ }\href {\doibase
  10.1103/PhysRevD.10.2491} {\bibfield  {journal} {\bibinfo  {journal} {Phys.
  Rev. D}\ }\textbf {\bibinfo {volume} {10}},\ \bibinfo {pages} {2491}
  (\bibinfo {year} {1974})}\BibitemShut {NoStop}%
\bibitem [{\citenamefont {Linde}(1977{\natexlab{b}})}]{Linde:1977mm}%
  \BibitemOpen
  \bibfield  {author} {\bibinfo {author} {\bibfnamefont {Andrei~D.}\
  \bibnamefont {Linde}},\ }\bibfield  {title} {\enquote {\bibinfo {title} {{On
  the Vacuum Instability and the Higgs Meson Mass}},}\ }\href {\doibase
  10.1016/0370-2693(77)90664-5} {\bibfield  {journal} {\bibinfo  {journal}
  {Phys. Lett. B}\ }\textbf {\bibinfo {volume} {70}},\ \bibinfo {pages}
  {306--308} (\bibinfo {year} {1977}{\natexlab{b}})}\BibitemShut {NoStop}%
\bibitem [{\citenamefont {Parisi}(1975)}]{Parisi:1975im}%
  \BibitemOpen
  \bibfield  {author} {\bibinfo {author} {\bibfnamefont {G.}~\bibnamefont
  {Parisi}},\ }\bibfield  {title} {\enquote {\bibinfo {title} {{The Theory of
  Nonrenormalizable Interactions. 1. The Large N Expansion}},}\ }\href
  {\doibase 10.1016/0550-3213(75)90624-0} {\bibfield  {journal} {\bibinfo
  {journal} {Nucl. Phys. B}\ }\textbf {\bibinfo {volume} {100}},\ \bibinfo
  {pages} {368--388} (\bibinfo {year} {1975})}\BibitemShut {NoStop}%
\bibitem [{\citenamefont {Itzhaki}\ \emph {et~al.}(1998)\citenamefont
  {Itzhaki}, \citenamefont {Maldacena}, \citenamefont {Sonnenschein},\ and\
  \citenamefont {Yankielowicz}}]{Itzhaki:1998dd}%
  \BibitemOpen
  \bibfield  {author} {\bibinfo {author} {\bibfnamefont {Nissan}\ \bibnamefont
  {Itzhaki}}, \bibinfo {author} {\bibfnamefont {Juan~Martin}\ \bibnamefont
  {Maldacena}}, \bibinfo {author} {\bibfnamefont {Jacob}\ \bibnamefont
  {Sonnenschein}}, \ and\ \bibinfo {author} {\bibfnamefont {Shimon}\
  \bibnamefont {Yankielowicz}},\ }\bibfield  {title} {\enquote {\bibinfo
  {title} {{Supergravity and the large N limit of theories with sixteen
  supercharges}},}\ }\href {\doibase 10.1103/PhysRevD.58.046004} {\bibfield
  {journal} {\bibinfo  {journal} {Phys. Rev. D}\ }\textbf {\bibinfo {volume}
  {58}},\ \bibinfo {pages} {046004} (\bibinfo {year} {1998})},\ \Eprint
  {http://arxiv.org/abs/hep-th/9802042} {arXiv:hep-th/9802042} \BibitemShut
  {NoStop}%
\bibitem [{\citenamefont {Klebanov}\ and\ \citenamefont
  {Polyakov}(2002)}]{Klebanov:2002ja}%
  \BibitemOpen
  \bibfield  {author} {\bibinfo {author} {\bibfnamefont {I.~R.}\ \bibnamefont
  {Klebanov}}\ and\ \bibinfo {author} {\bibfnamefont {A.~M.}\ \bibnamefont
  {Polyakov}},\ }\bibfield  {title} {\enquote {\bibinfo {title} {{AdS dual of
  the critical O(N) vector model}},}\ }\href {\doibase
  10.1016/S0370-2693(02)02980-5} {\bibfield  {journal} {\bibinfo  {journal}
  {Phys. Lett. B}\ }\textbf {\bibinfo {volume} {550}},\ \bibinfo {pages}
  {213--219} (\bibinfo {year} {2002})},\ \Eprint
  {http://arxiv.org/abs/hep-th/0210114} {arXiv:hep-th/0210114} \BibitemShut
  {NoStop}%
\bibitem [{\citenamefont
  {Romatschke}(2019{\natexlab{b}})}]{Romatschke:2019ybu}%
  \BibitemOpen
  \bibfield  {author} {\bibinfo {author} {\bibfnamefont {Paul}\ \bibnamefont
  {Romatschke}},\ }\bibfield  {title} {\enquote {\bibinfo {title}
  {{Finite-Temperature Conformal Field Theory Results for All Couplings: O(N)
  Model in 2+1 Dimensions}},}\ }\href {\doibase 10.1103/PhysRevLett.122.231603}
  {\bibfield  {journal} {\bibinfo  {journal} {Phys. Rev. Lett.}\ }\textbf
  {\bibinfo {volume} {122}},\ \bibinfo {pages} {231603} (\bibinfo {year}
  {2019}{\natexlab{b}})},\ \bibinfo {note} {[Erratum: Phys.Rev.Lett. 123,
  209901 (2019)]},\ \Eprint {http://arxiv.org/abs/1904.09995} {arXiv:1904.09995
  [hep-th]} \BibitemShut {NoStop}%
\bibitem [{\citenamefont {Romatschke}(2020)}]{Romatschke:2019wxc}%
  \BibitemOpen
  \bibfield  {author} {\bibinfo {author} {\bibfnamefont {Paul}\ \bibnamefont
  {Romatschke}},\ }\bibfield  {title} {\enquote {\bibinfo {title} {{Simple
  non-perturbative resummation schemes beyond mean-field II: Thermodynamics of
  scalar $\phi^4$ theory in 1 + 1 dimensions at arbitrary coupling}},}\ }\href
  {\doibase 10.1142/S0217732320500546} {\bibfield  {journal} {\bibinfo
  {journal} {Mod. Phys. Lett. A}\ }\textbf {\bibinfo {volume} {35}},\ \bibinfo
  {pages} {2050054} (\bibinfo {year} {2020})},\ \Eprint
  {http://arxiv.org/abs/1903.09661} {arXiv:1903.09661 [hep-th]} \BibitemShut
  {NoStop}%
\bibitem [{\citenamefont {Weiner}\ and\ \citenamefont
  {Romatschke}(2023)}]{Weiner:2022kgx}%
  \BibitemOpen
  \bibfield  {author} {\bibinfo {author} {\bibfnamefont {Max}\ \bibnamefont
  {Weiner}}\ and\ \bibinfo {author} {\bibfnamefont {Paul}\ \bibnamefont
  {Romatschke}},\ }\bibfield  {title} {\enquote {\bibinfo {title} {{Determining
  all thermodynamic transport coefficients for an interacting large N quantum
  field theory}},}\ }\href {\doibase 10.1007/JHEP01(2023)046} {\bibfield
  {journal} {\bibinfo  {journal} {JHEP}\ }\textbf {\bibinfo {volume} {01}},\
  \bibinfo {pages} {046} (\bibinfo {year} {2023})},\ \Eprint
  {http://arxiv.org/abs/2208.10502} {arXiv:2208.10502 [hep-th]} \BibitemShut
  {NoStop}%
\bibitem [{\citenamefont {Cea}(2020)}]{Cea:2020lea}%
  \BibitemOpen
  \bibfield  {author} {\bibinfo {author} {\bibfnamefont {Paolo}\ \bibnamefont
  {Cea}},\ }\bibfield  {title} {\enquote {\bibinfo {title} {{The Higgs
  condensate as a quantum liquid}},}\ }\href {\doibase
  10.1007/s10773-020-04589-9} {\bibfield  {journal} {\bibinfo  {journal} {Int.
  J. Theor. Phys.}\ }\textbf {\bibinfo {volume} {59}},\ \bibinfo {pages}
  {3310--3323} (\bibinfo {year} {2020})},\ \Eprint
  {http://arxiv.org/abs/2003.07188} {arXiv:2003.07188 [hep-ph]} \BibitemShut
  {NoStop}%
\end{thebibliography}%
\end{document}